\documentclass[
aps,prd,
12pt,%10pt
nopreprintnumbers,
showpacs,
eqsecnum,
nofootinbib
]{revtex4-1}

\usepackage{graphicx}
\usepackage{amssymb}

\begin{document}

\title{Discrete time heat kernel and UV modif{}ied propagators with Dimensional
Deconstruction}
\author{Nahomi Kan}\email[]{kan@gifu-nct.ac.jp}
\affiliation{National Institute of Technology, Gifu College,
Motosu-shi, Gifu 501-0495, Japan}
%\author{Takuma Aoyama}\email[]{b014vbv@yamaguchi-u.ac.jp}
%\affiliation{
%Graduate School of Sciences and Technology for Innovation, Yamaguchi
%University, Yamaguchi-shi, Yamaguchi 753--8512, Japan}
\author{Kiyoshi Shiraishi}\email[]{shiraish@yamaguchi-u.ac.jp}
\affiliation{
Graduate School of Sciences and Technology for Innovation, Yamaguchi
University, Yamaguchi-shi, Yamaguchi 753--8512, Japan}
\date{\today}
%\date{}

\begin{abstract}
We revisit the dimensionally
deconstructed scalar quantum electrodynamics and consider the (Euclidean)
propagator of the scalar field in the model. Although we have previously
investigated the one-loop effect in this model by obtaining the usual heat
kernel trace, we adopt discrete proper-time heat kernels in this paper and aim to
construct the modified propagator, which has improved behaviors in the ultraviolet
region, by changing the range of sum of the discrete heat kernels.
\end{abstract}

%\preprint{}

\pacs{%
%02.10.Ox, %%%Combinatorics; graph theory
%02.20.Sv, %Lie algebra of Lie groups
%02.30.Hq, %Ordinary differential equations
%02.30.Ik, %Integrable systems
%02.30.Jr, %Partial differential equations
%02.40.Gh, %Noncommutative geometry
%03.65.-w, %Quantum mechanics
%03.65.Sq, %Semiclassical theories in quantum mechanics
03.70.+k, %Theory of quantized fields
%04.20.-q, %%%Classical general relativity
%04.20.Fy, %%Canonical formalism, Lagrangians, and variational principles
%04.20.Jb, %%Exact solutions
%04.25.-g, %Approximation
%04.25.Nx, %%%Post-Newtonian approximation; perturbation theory; related
%approximations
%04.40.-b, %Self-Gravitating systems
%04.40.Nr, %%Einstein-Maxwell spacetime
%04.50.-h, %%%%%Higher-dimensional gravity and other theories of gravity 
%04.50.Cd, %Kaluza--Klein theories 
%04.50.Gh, %Higher-dimensional black holes, black strings, 
%and related objects 
%04.50.Kd, %%%Modified theories of gravity 
%04.60.-m, %%Quantum gravity
%04.60.Kz, %%Lower dimensional models; minisuperspace models
%04.60.Rt, %Topologically massive gravity
%04.62.+v, %Quantum fields in curved spacetime
%04.65.+e, %Supergravity
%04.70.Bw, %%%Classical black holes
%05.30.-d, %Quantum statistical mechanics
%05.30.Jp, %Boson systems
%11.10.-z, %%%Field theory
%11.10.Ef, %%%Lagrangian and Hamiltonian approach
11.10.Lm, %%%Nonlinear or nonlocal theories and models 
%11.10.Nx, %%%Noncommutative field theory 
11.10.Kk, %%%Field theories in dimensions other than four
%11.10.Wx, %Finite-temperature field theory 
%11.25.-w, %Strings and branes
%11.25.Mj, %%Compactification and four-dimensional models
%11.27.+d %%Extended classical solutions; cosmic strings, 
%domain walls, texture 
%11.30.-j, %Symmetry and conservation laws
%11.30.Pb, %Supersymmetry
%12.60.-i, %Models beyond the standard model
%12.90.+b, %Miscellaneous theoretical ideas and models
%45.20.Jj, %Lagrangian and Hamiltonian mechanics
%95.35.+d, %Dark matter
%95.36.+x, %Dark energy
%98.80.-k, %%%Cosmology 
%98.80.Cq, %%%%%Particle-theory and field-theory models of the early
%Universe  
%98.80.Dr, %Relativistic cosmology 
%98.80.Qc, %Quantum cosmology
%98.80.Jk% %%Mathematical and relativistic aspects of cosmology
.}

\maketitle

%%%%%%%%%%%%%%%%%%%%%%%%%%%%%%%%%%%%%%%%%%%%%%%%%%%%%%%%%%%%%%%%%%%%%%%%%%%
%%%%%%%%%%%%%%%%%%%%%%%%%%%%%%%%%%%%%%%%%%%%%%%%%%%%%%%%%%%%%%%%%%%%%%%%%%%
%%%%%%%%%%%%%%%%%%%%%%%%%%%%%%%%%%%%%%%%%%%%%%%%%%%%%%%%%%%%%%%%%%%%%%%%%%%
\section{Introduction}
\label{introduction}
%%%%%%%%%%%%%%%%%%%%%%%%%%%%%%%%%%%%%%%%%%%%%%%%%%%%%%%%%%%%%%%%%%%%%%%%%%%
%%%%%%%%%%%%%%%%%%%%%%%%%%%%%%%%%%%%%%%%%%%%%%%%%%%%%%%%%%%%%%%%%%%%%%%%%%%
%%%%%%%%%%%%%%%%%%%%%%%%%%%%%%%%%%%%%%%%%%%%%%%%%%%%%%%%%%%%%%%%%%%%%%%%%%%

In recent years, various ideas have been discussed to consider \cite{%
Pad1,Pad2,SSP,SP,KSSP,Pad3,Pad4,Pad5,Pad6,Pad7,Pad8,Pad9,Nicolini,
AD,ABM,AL,Siegel,KKSW1,KKSW2,Mondal} how to take changes in the
Schwinger's proper-time parameter of the heat kernel \cite{Vassilevich,Camporesi}
for a certain Laplace operator 
as a
way to improve ultraviolet (UV) behavior in quantum field theory. In our previous
paper \cite{KKSW2}, we proposed a method to %systematically 
derive 
propagators (Green's functions) with moderate behavior at an
infinitesimally short distance by discretizing the proper-time parameter and
adjusting the range of summation. The modified propagator obtained with this
prescription is corresponding to the one in the theory of the original canonical
field action modified by adding higher-order derivative terms \cite{KKSW2}.

For the simplest example of the self-interacting canonical scalar field, consider
the $\lambda\varphi^4$ theory. The scalar one-loop effect gives the
radiative correction
$\delta m^2\sim\lambda\langle\varphi^2\rangle$ (with zero momentum transfer).
This suffers from a $(D-2)$-th order divergence in $D$-dimensional
spacetime, since
$\langle\varphi^2\rangle\sim G(x,x)\sim\int d^Dp/p^2$, where $G(x,x')$ is the
propagator of the scalar field, and the Fourier-transformed propagator
$\tilde{G}(p^2)$ then behaves as $\sim 1/p^2$ at high energy. Phenomenologically,
this fact is known as an origin of the hierarchy problem in the standard model of
particle physics with $D=4$.%
\footnote{Needless to say, loop effects of fermions such as heavy quarks are also
important in the hierarchy problem. We will discuss them in the last section.}
 The way to obtain softer behavior of
$\tilde{G}(p^2)\sim 1/p^4$ or more at high energy has been studied for some time
until now \cite{LW,GOW,CL}.

On the other hand, many authors have also considered the direction of assuming 
the discretized background spacetime (motivated by a way of thinking about
quantum gravitational consideration \cite{tHooft}). By the way, in a limited
sense, discretization is reminiscent of dimensional deconstruction (DD)
\cite{ACG,HPW,HL}. The idea of DD has even been
incorporated into phenomenological models and variously explored. 
Suppose a number of copies of a four-dimensional theory and linking pairs of these
individual sites in the theory space.
The resulting whole theory mimics a higher-dimensional theory. This is an
attempt to introduce the discrete extra space into the theory, and
so interesting characteristics of higher dimensional theory can be inherited
by four dimensional theory.
We should note, however, that high energy behavior of the theory becomes even worse
because the extra dimensional contributions are summed up to obtain a four
dimensional effective theory.

In the present paper, we study the UV modified propagators in the
deconstructed model by using the discrete time heat kernel. Concretely, we find the
Euclidean propagator $G(x,x')$ of the complex scalar fields $\phi_\nu$ and its
coincidence limit $G(x,x)$, which is proportional to the one-loop vacuum
polarization
$\langle\phi^\dagger_\nu\phi_\nu\rangle$, of the dimensionally deconstructed
$(D+1)$-dimensional scalar quantum electrodynamics (QED). Here we assume the
presence of the background of pseudo-Nambu--Goldstone boson (PNGB) field
\cite{HPW,HL} (which corresponds to the constant $U(1)$ gauge field in a
higher-dimensional theory) in one discrete extra dimensional direction.

The model treated here is the same as the one studied in Ref.~\cite{KSS}
using the usual heat kernel trace method. In the study, we have a heat kernel trace
that employs the graph Laplacian associated to a cycle graph as a part of the
Laplace operator, known from spectral graph theory,%
\footnote{For the spectral graph theory, see
Refs.~\cite{Mohar1,Mohar2,Mohar3,Merris}.}
 which coincides
with  that obtained from the heat kernel discussed more generally later in
Refs.~\cite{CJK1,CJK2,Dowker}.
It is known that the DD model using the cycle graph has a continuous limit, which
yields the Kaluza--Klein model with an extra dimension $S^1$.
It should be noted here that, as we have already pointed out in
Ref.~\cite{KKSW2}, even if the Laplace operator is in the form of a direct sum,
the heat kernel cannot be expressed in a form of a direct product
when the proper time is discrete.
Therefore, it is significant to study the mathematical properties of the discrete
time heat kernel, albeit for a simple model. Fortunately, discrete time heat
kernel for a certain class of graph Laplacians has recently been discussed in
Ref.~\cite{CHJSV}, so we can manage to apply it to our calculations.

Although the model considered in this paper is the simplest one, there is a future
goal to develop this toy model into various field theories with higher symmetries
to explore continuous and discrete versions of the Hosotani mechanism
\cite{Hosotani} with UV modification. At the same time, extension from DD to
models using various graph Laplacians will also come into future view.
As a natural extension of DD, we can study the discretization of
our real spacetime in a similar method, which we will link to future research
and work by other authors  \cite{tHooft}.

The structure of this paper is as follows. In Section \ref{sec2}, we review the
model setup, the derivation of the usual heat kernel, and the calculations
leading to the propagator. They are necessary for the
comparison with those obtained from the discrete time heat kernel later. Section
\ref{sec3} introduces two types of discrete time heat kernels and discusses how to
soften or moderate the UV divergence in the short range behavior of the (Euclidean)
propagator by changing the sum of kernels.  The final section is devoted to
summary and future prospects.

%%%%%%%%%%%%%%%%%%%%%%%%%%%%%%%%%%%%%%%%%%%%%%%%%%%%%%%%%%%%%%%%%%%%%%%%%%%
%%%%%%%%%%%%%%%%%%%%%%%%%%%%%%%%%%%%%%%%%%%%%%%%%%%%%%%%%%%%%%%%%%%%%%%%%%%
%%%%%%%%%%%%%%%%%%%%%%%%%%%%%%%%%%%%%%%%%%%%%%%%%%%%%%%%%%%%%%%%%%%%%%%%%%%
\section{(continuous time) heat kernel for DD with a cycle graph}
\label{sec2}
%%%%%%%%%%%%%%%%%%%%%%%%%%%%%%%%%%%%%%%%%%%%%%%%%%%%%%%%%%%%%%%%%%%%%%%%%%%
%%%%%%%%%%%%%%%%%%%%%%%%%%%%%%%%%%%%%%%%%%%%%%%%%%%%%%%%%%%%%%%%%%%%%%%%%%%
%%%%%%%%%%%%%%%%%%%%%%%%%%%%%%%%%%%%%%%%%%%%%%%%%%%%%%%%%%%%%%%%%%%%%%%%%%%

We revisit the deconstructed massless scalar QED model introduced in
Ref.~\cite{KSS}. Its action of the scalar field sector is expressed as follows:
\begin{equation}
S=-\sum_{\nu=1}^N\sum_{\nu'=1}^N\int
d^Dx\,\phi_\nu^\dagger(x)\Bigl[-I_{\nu\nu'}\Box+f^2\Delta_{\nu\nu'}(C_N,\chi)\Bigr]
\phi_{\nu'}(x)\,,
\label{2.1}
\end{equation}
where $I_{\nu\nu'}$ denotes the $N\times N$ identity matrix and
 the Hermitian matrix $\Delta(C_N,\chi)$ is given by
\begin{equation}
\Delta(C_N,\chi)\equiv\left(
\begin{array}{cccccc}
2&-e^{i\chi}&0&\cdots& &-e^{-i\chi}\\
-e^{-i\chi}&2&-e^{i\chi}&\cdots& &0\\
0&-e^{-i\chi}&2&\cdots& &0\\
\vdots&\vdots&\vdots&\ddots& &\vdots\\
 & & & &2&-e^{i\chi}\\
-e^{i\chi}&0&0&\cdots&-e^{-i\chi}&2
\end{array}
\right)\,,
\label{mtrx}
\end{equation}
and $f$ is a constant with dimension of mass.
We consider the $D$-dimensional Euclidean space.
The d'Alembert operator
$\Box=\sum_{j=0}^{D-1}\partial^j\partial_j$ in (\ref{2.1}) acts on scalar fields.
The label of the fields are considered as periodic modulo $N$, e.g.,
$\phi_{N+1}\equiv\phi_1$, $\phi_0\equiv\phi_N$, and so on. 
The matrix $\Delta(C_N,0)$ is the graph Laplacian for a cycle graph $C_N$
\cite{Mohar1,Mohar2,Mohar3,Merris}, whose
$N$ vertices form a discrete circle.
The constant $\chi$ stands for the `twist' factor, which comes from the constant
background PNGB field corresponding to the background $U(1)$ gauge field in the
extra dimensions if the continuous limit is taken. We omit the background gauge
field in the flat large dimensions in the present analysis.

Here, we introduce a heat kernel $K_\nu(x,x';s)$ that satisfies the
equation
\begin{equation}
\sum_{\nu'=1}^N\left[I_{\nu\nu'}\frac{\partial}{\partial
s}+\Bigl(-I_{\nu\nu'}\Box_x
+f^2\Delta_{\nu\nu'}(C_N,\chi)\Bigr)\right]K_{\nu'}(x,x';s)=0\,,
\end{equation}
subject to the initial condition $\lim_{s\rightarrow
0}K_0(x,x';s)=\delta(x,x')$ and $\lim_{s\rightarrow
0}K_\nu(x,x';s)=0$ for $\nu\ne 0$. 
The d'Alembert operator $\Box_x$ acts on the coordinate $x$.
In our present model, the heat kernel can be written in the form%
\footnote{Obviously, we use $\nu-\nu'$ instead of $\nu$ for an arbitrary pair of
the scalar fields on the sites labeled by $\nu$ and $\nu'$.}
\begin{equation}
K_\nu(x,x';s)=\int\frac{d^Dp}{(2\pi)^D} \tilde{K}_\nu(p^2;s) e^{ip\cdot
(x-x')}\,,
\label{he1}
\end{equation}
where we notice that $\tilde{K}_\nu(p^2;s)$ is a function of
$p^2=\sum_{j=1}^{D-1}p_jp^j$, for the homogeneity and isotropy of the Euclidean
space $\mathbf{R}^D$.
Then, the heat equation (\ref{he1}) reduces to
\begin{equation}
{\partial_s}
\tilde{K}_\nu(p^2;s)=f^2\Bigl[e^{i\chi}\tilde{K}_{\nu+1}(p^2;s)-
(2+p^2/f^2)\tilde{K}_\nu(p^2;s)+e^{-i\chi}\tilde{K}_{\nu-1}(p^2;s)\Bigr]\,,
\label{he2}
\end{equation}
with the condition $\lim_{s\rightarrow
0}\tilde{K}_0(p^2;s)=1$ and $\lim_{s\rightarrow
0}\tilde{K}_\nu(p^2;s)=0$ for $\nu\ne 0$.
Here, the abbreviation $\partial_s\equiv\frac{\partial}{\partial s}$ has been used.

The solution for (\ref{he2}) can be expressed by using the modified Bessel function
$I_\nu(z)$ \cite{GR}
\begin{equation}
I_\nu(z)=\left(\frac{z}{2}\right)^\nu\sum_{n=0}^\infty\frac{(z/2)^{2n}}{n!\Gamma(\nu+n+1)}\,.
\end{equation}
Note that $I_0(0)=1$, $I_\nu(0)=0$ for $\nu\ne 0$, and $I_{-\nu}(z)=I_\nu(z)$ if
$\nu\in\mathbf{Z}\equiv
\{\dots, -2, -1, 0, 1, 2, \dots\}$. We should also notice that
\begin{equation}
\partial_zI_\nu(z)=\frac{1}{2}[I_{\nu+1}(z)+I_{\nu-1}(z)]\,.
\end{equation}

A solution for the heat equation is found to be \cite{CJK1,CJK2,Dowker}
\begin{equation}
e^{-i\nu\chi}e^{-(p^2+2f^2)s}I_\nu(2f^2s)\,,
\end{equation}
which becomes unity if $\nu=s=0$ and vanishes for the case with  $s=0$ and $\nu\ne
0$.
For our present model, the kernel should be periodic
such as $K_{\nu+N}(p^2;s)=K_\nu(p^2;s)$. Therefore, the solution of the heat
equation subject to the boundary conditions turns out to be
\begin{equation}
\tilde{K}_\nu(p^2;s)=e^{-(p^2+2f^2)s}\sum_{q=-\infty}^\infty
e^{-i(\nu+qN)\chi}I_{\nu+qN}(2f^2s)\,.
\end{equation}

The `genuine' trace of the heat kernel is written as
\begin{eqnarray}
N\int d^Dx\, K_0(x,x;s)&=&NV\int\frac{d^Dp}{(2\pi)^D}\tilde{K}_0(p^2;s)\nonumber \\
&=&\frac{NV}{(4\pi)^{D/2}s^{D/2}}e^{-2f^2s}\sum_{q=-\infty}^\infty
\cos(qN\chi)I_{qN}(2f^2s)\,.
\end{eqnarray}
Here, $V$ denotes the volume of the Euclidean spacetime $\int d^Dx$,
which is often omitted in the works on the heat kernel trace in quantum field
theory. Similarly, the factor $N$ is regarded as the `volume' of the discrete
circle, or simply recognized as the `symmetry factor' for $N$ scalar fields.

In the paper Ref.~\cite{KSS}, we sought the heat kernel trace from the beginning.
Here we briefly describe it.
The eigenvalues of the matrix (\ref{mtrx}) are $4\sin^2\left(\frac{\pi
k}{N}+\frac{\chi}{2}\right)$ ($k=0, 1, \dots, N-1$).
Then, utilizing the mathematical formula \cite{GR}, we find
\begin{eqnarray}
\sum_{k=0}^{N-1} \exp\left[-4f^2\sin^2\left(\frac{\pi k}{N}+\frac{\chi}{2}\right)s
\right]&=&e^{-2f^2 s}\sum_{p=0}^{N-1}\sum_{\ell=-\infty}^\infty\cos
\left[\ell\left(\frac{2\pi p}{N}+{\chi}\right)\right]I_\ell(2f^2 s)\nonumber \\
&=&N e^{-2f^2 s}\sum_{q=-\infty}^{\infty}\cos
\left(qN{\chi}\right)I_{qN}(2f^2 s)\,,
\end{eqnarray}
so, the heat kernel trace has been certainly reproduced.

Now, we consider the `partial trace', i.e., the trace only on the label of
scalar fields (or equivalently, vertices). This operation corresponds to
`integrating out the extra discrete dimensions'%
\footnote{Consequently, the one-loop effect which comes from the extra discrete
circle is included.}  and creating a physical $D$-dimensional perspective.
Accordingly, the partially-traced Fourier transform of the propagator is
\begin{eqnarray}
& &\tilde{G}(p^2)\equiv N\int_0^\infty \tilde{K}_0(p^2;s)\, ds
=N\int_0^\infty e^{-(p^2+2f^2)s}\sum_{q=-\infty}^\infty
\cos(qN\chi)I_{qN}(2f^2s) ds\nonumber \\
%&=&N(p^2+2f^2)^{-1}\sum_{q=-\infty}^\infty\cos(qN\chi)
%\left(\frac{f^2}{p^2+2f^2}\right)^{qN}
%F\Bigl(\frac{qN+1}{2},\frac{qN+2}{2};qN+1;\frac{4f^4}{(p^2+2f^2)^2}\Bigr)
%\nonumber\\
%&=&\frac{N}{p^2}\sum_{q=-\infty}^\infty\cos(qN\chi)
%\left(\frac{f^2}{p^2}\right)^{|q|N}
%F\Bigl(|q|N+1,|q|N+\frac{1}{2};2|q|N+1;-\frac{4f^2}{p^2}\Bigr)\nonumber \\
&=&\frac{N}{\sqrt{p^2(4f^2+p^2)}}\sum_{q=-\infty}^\infty\cos(qN\chi)
\left(\frac{2f^2}{2f^2+p^2+\sqrt{p^2(4f^2+p^2)}}\right)^{|q|N}\nonumber \\
&=&\frac{N/f^2}{2\sinh\beta}
\frac{\sinh(N\beta)}{\cosh(N\beta)-\cos(N\chi)}\,,
\label{intg}
\end{eqnarray}
where, in the last line, the new parameter
\begin{equation}
\frac{p^2}{f^2}\equiv 4\sinh^2\frac{\beta}{2}
\sim\left\{
\begin{array}{cc}
\beta^2 & (p^2/f^2\ll 1)\\
e^\beta & (p^2/f^2\gg 1)
\end{array}
\right.\,,
\end{equation}
has been used.

As a special but familiar case, for $\chi=0$, we find
\begin{equation}
\tilde{G}(p^2)\Bigl|_{\chi=0}=\frac{N/f^2}{2\sinh\beta}
\frac{\cosh(N\beta/2)}{\sinh(N\beta/2)}\,,
\end{equation}
and further if $\beta$ is small,%
\footnote{Of course, as a trivial check, we also observe 
$\tilde{G}(p^2)|_{\chi=0}=1/p^2$ if $N=1$.}
$\tilde{G}(p^2)|_{\chi=0}\approx
1/(f^2\beta^2)\approx {1}/{p^2}$, that is the usual Fourier-transformed propagator,
also known as the propagator in the momentum space.%
\footnote{Note that the addition of a common mass $m$ to scalar fields leads to
replacing $p^2\rightarrow p^2+m^2$ in the propagator in
momentum space.} Then, the propagator in the 
$D$-dimensional spacetime coordinates,
\begin{equation}
G(x,x')=\int\frac{d^Dp}{(2\pi)^D}\,\tilde{G}(p^2)\, e^{ip\cdot(x-x')}\,,
\end{equation}
is proportional to $1/r^{D-2}$,
where
$r=|x-x'|$. On the other hand, if $p^2/f^2\gg 1$, $\tilde{G}(p^2)|_{\chi=0}\approx
N/p^2$, as expected for finite $N$. What we have to be careful about is when
$N/f\equiv L$ is fixed and $N$ and $f$ approaches infinity, then
$\tilde{G}(p^2)|_{\chi=0}\approx
 L/(2p)$ for $pL\gg 1$. Therefore, in this case, $G(x,x')\propto L/r^{D-1}$ for a
small $r$, whose behavior corresponds to the $(D+1)$ dimensional one.

Next, let us see the divergence behavior of
$G(x,x)\sim\langle\phi^\dagger_\nu\phi_\nu\rangle$ for $D=4$. 
If we use the cutoff momentum $\Lambda$, we find
\begin{eqnarray}
G(x,x)&=&\frac{2\pi^2 N}{(2\pi)^4}\int_0^\Lambda  \tilde{G}(p^2)\,p^3 dp\nonumber
\\ &=&\frac{N}{16\pi^2}\left[\sqrt{\Lambda^2(\Lambda^2+4f^2)}-4f^2\ln\left(
\frac{\Lambda}{2f}+\sqrt{1+\frac{\Lambda^2}{4f^2}}\right)\right]\nonumber \\
& &-\sum_{q=1}^\infty\frac{2}{8\pi^3q(q^2N^2-1)}\cos(qN\chi)
\Biggl[\left(q^2N^2\Lambda^2+qN\sqrt{\Lambda^2(\Lambda^2+4f^2)}+2f^2\right)\nonumber
\\ & &\times
\left(\frac{2f^2}{\Lambda^2+2f^2+\sqrt{\Lambda^2(\Lambda^2+4f^2)}}\right)^{Nq}
-2f^2\Biggr]\,.
\end{eqnarray}
The first term in the last line, which corresponds to the $q=0$ term, apparently
diverges quadratically when $\Lambda\rightarrow\infty$.
This divergent term is independent of $\chi$, the background PNGB field.
For $N\ge 2$, the remaining terms converge when $\Lambda\rightarrow\infty$,
and become
\begin{equation}
\sum_{q=1}^\infty\frac{f^2}{2\pi^3q(q^2N^2-1)}\cos(qN\chi)\,.
\end{equation}

Consequently, we find that $G(x,x)$ has the quadratic divergence ($\sim\Lambda^2$)
in four dimensions, and the divergent term is independent of the background field
$\chi$ in the present model.

Finally, in the remainder of this section we see the further correspondence with
already known results. This will also serve as a check on our calculations. Because
$\tilde{G}(p^2)$ can be regarded as the trace of an inverse matrix
$A(p^2)^{-1}$, where
\begin{equation}
A(p^2)=f^2\left[(p^2/f^2)I+\Delta(C_N,\chi)\right]\,,
\end{equation}
the integral connects the propagator $\tilde{G}(p^2)$ and the determinant of
$A(p^2)$ as follows.
\begin{eqnarray}
\ln\Biggl[\frac{\det A(p^2)}{\det A(\Lambda^2)}\Biggr]
&=&\int_{\Lambda^2}^{p^2}\tilde{G}(\mu^2)\,d\mu^2
=N\int_{\lambda}^{\beta}
\frac{\sinh(Nz)}{\cosh(Nz)-\cos(N\chi)}dz\nonumber \\
&=&\ln\Biggl[\frac{\cosh(N\beta)-\cos(N\chi)}{\cosh(N\lambda)-\cos(N\chi)}\Biggr]
=\ln\Biggl[
\frac{\sinh^2(N\beta/2)+\sin^2(N\chi/2)}{\sinh^2(N\lambda/2)+\sin^2(N\chi/2)}
\Biggr]\,,
\end{eqnarray}
where $\Lambda$ is a constant and
$4\sinh^2\frac{\lambda}{2}=\frac{\Lambda^2}{f^2}$.
This result is consistent with the calculation of the determinant
found in Ref.~\cite{KS}.
Incidentally, this determinant can be used to calculate the one-loop vacuum
energy.%
\footnote{The method to obtain the one-loop effective action from the
propagator has been well-known, for example, see Refs.~\cite{CR,DC}.}%
\footnote{The one-loop contribution of the vector field is neglected in this
time.} Of course, for $D=4$, there is a quartic divergence, but since the part
depending on
$\chi$ is finite, the four dimensional case gives the well-known form of the
effective potential for
$\chi$:\cite{ACG,HPW,HL,KSS}
\begin{equation}
V(\chi)=-\frac{3f^4}{2\pi^2}\sum_{q=1}^{\infty}
\frac{\cos(qN\chi)}{q(q^2N^2-1)(q^2N^2-4)}\,.\quad(N\ge 3)
\end{equation}

In the next section, we will consider the discrete time heat kernel.

%%%%%%%%%%%%%%%%%%%%%%%%%%%%%%%%%%%%%%%%%%%%%%%%%%%%%%%%%%%%%%%%%%%%%%%%%%%

%%%%%%%%%%%%%%%%%%%%%%%%%%%%%%%%%%%%%%%%%%%%%%%%%%%%%%%%%%%%%%%%%%%%%%%%%%%
%%%%%%%%%%%%%%%%%%%%%%%%%%%%%%%%%%%%%%%%%%%%%%%%%%%%%%%%%%%%%%%%%%%%%%%%%%%
%%%%%%%%%%%%%%%%%%%%%%%%%%%%%%%%%%%%%%%%%%%%%%%%%%%%%%%%%%%%%%%%%%%%%%%%%%%
\section{discrete time heat kernel for DD with a cycle graph}
\label{sec3}
%%%%%%%%%%%%%%%%%%%%%%%%%%%%%%%%%%%%%%%%%%%%%%%%%%%%%%%%%%%%%%%%%%%%%%%%%%%
%%%%%%%%%%%%%%%%%%%%%%%%%%%%%%%%%%%%%%%%%%%%%%%%%%%%%%%%%%%%%%%%%%%%%%%%%%%
%%%%%%%%%%%%%%%%%%%%%%%%%%%%%%%%%%%%%%%%%%%%%%%%%%%%%%%%%%%%%%%%%%%%%%%%%%%

The discrete time heat kernels for the $D$-dimensional canonical scalar model was
discussed in Ref.~\cite{KKSW2}. The discrete time heat kernels for the graph
Laplacian are recently dealt with in Ref.~\cite{CHJSV}. Here we extend the
technique to the case of the DD model investigated so far. 

There are two types of the difference operator:
The forward difference operator $\Delta$ is defined by  
\begin{equation}
\Delta f(t)\equiv f(t+1)-f(t)\,,
\end{equation}
while the backward difference operator $\nabla$ is defined by
\begin{equation}
\nabla f(t)\equiv f(t)-f(t-1)\,.
\end{equation}

In the two subsections below, we will obtain the heat kernel as a solution of the
equation using each type of difference, and show the construction of the
momentum-space propagator. After that, the UV modification is studied in the third
subsection.

%%%%%%%%%%%%%%%%%%%%%%%%%%%%%%%%%%%%%%%%%%%%%%%%%%%%%%%%%%%%%%%%%%%%%%%%%%%
%%%%%%%%%%%%%%%%%%%%%%%%%%%%%%%%%%%%%%%%%%%%%%%%%%%%%%%%%%%%%%%%%%%%%%%%%%%
\subsection{The solution of the forward difference equation}
%%%%%%%%%%%%%%%%%%%%%%%%%%%%%%%%%%%%%%%%%%%%%%%%%%%%%%%%%%%%%%%%%%%%%%%%%%%
%%%%%%%%%%%%%%%%%%%%%%%%%%%%%%%%%%%%%%%%%%%%%%%%%%%%%%%%%%%%%%%%%%%%%%%%%%%

In this subsection, we consider the discrete time heat kernel defined as the
unique solution of the forward difference equation 
\begin{equation}
\Delta\tilde{K}_\nu(p^2;t)=\epsilon
f^2\Bigl[e^{i\chi}\tilde{K}_{\nu+1}(p^2;t)-
(2+p^2/f^2)\tilde{K}_\nu(p^2;t)+e^{-i\chi}\tilde{K}_{\nu-1}(p^2;t)\Bigr]\,,
\label{dhe}
\end{equation}
with the condition $\tilde{K}_0(p^2;0)=1$ and $\tilde{K}_\nu(p^2;0)=0$ for $\nu\ne
0$.
Compared with the differential equation (\ref{he2}), we find that the parameter $s$
corresponds to $\epsilon t$ ($t\in\mathbf{N}_0\equiv\{0, 1, 2, \dots\}$) and that
the differential equation (\ref{he2}) is recovered by the continuum limit,  since
$\frac{1}{\epsilon}\Delta\rightarrow \frac{\partial}{\partial s}$, if
$\epsilon\rightarrow 0$.

The solution of (\ref{dhe}) can be expressed by using
the discrete modified Bessel function $I^c_\nu(t)$ \cite{CHJSV,BC,Slavik},
\begin{eqnarray}
I^c_\nu(t)&\equiv&\frac{(-c/2)^\nu \Gamma(-t+\nu)}{\nu!\, \Gamma(-t)}F\Bigl(
\frac{\nu-t}{2},\frac{\nu-t+1}{2};\nu+1;c^2\Bigr)\nonumber \\
&=&\sum_{n=0}^\infty\frac{\Gamma(t+1)(c/2)^{2n+\nu}}{n!\,\Gamma(t-2n-\nu+1)\Gamma(\nu+n+1)}
\,,
\end{eqnarray}
where $F(\alpha,\beta;\gamma;z)$ is the Gauss' hypergeometric function.
Note that $I^c_{-\nu}(z)=I^c_{\nu}(z)$ for $\nu\in\mathbf{Z}$.

%%%%

Note also that we can express $I^c_\nu(t)$ as the form
\begin{equation}
I^c_\nu(t)=\frac{(c/2)^\nu t^{\underline{\nu}}}{\nu!}F\Bigl(
\frac{\nu-t}{2},\frac{\nu-t+1}{2};\nu+1;c^2\Bigr)\,.
\end{equation}
Here the falling power $t^{\underline{\nu}}$ is defined as
\begin{equation}
t^{\underline{\nu}}\equiv(-1)^\nu(-t)_\nu\,,
\end{equation}
where the Pochhammer symbol means
\begin{equation}
(x)_k\equiv x(x+1)\cdots(x+k-1)=\Gamma(x+k)/\Gamma(x)\,.
\end{equation}
One can see that $\Delta\, t^{\underline{n}}=n\,t^{\underline{n-1}}$.
It has been known that $I^1_\nu(t)$ is obtained by replacing $t^k$ in 
the Maclaurin series of $I_\nu(t)$ by $t^{\underline{k}}$ \cite{BC}.

The key property of $I^c_\nu(t)$ is that
\begin{equation}
\Delta
I^c_\nu(t)=I^c_\nu(t+1)-I^c_\nu(t)=\frac{c}{2}[I^c_{\nu+1}(t)+I^c_{\nu-1}(t)]\,.
\end{equation}
Therefore, the solution of the discrete heat equation (\ref{dhe}) can be written as
\begin{equation}
\tilde{K}_\nu(p^2;t)=\sum_{q=-\infty}^\infty e^{-i(\nu+qN)\chi} a^t
I^b_{\nu+qN}(t)\,,
\end{equation}
where
\begin{equation}
a\equiv 1-\epsilon f^2(2+p^2/f^2)\quad\mbox{and}\quad
b\equiv 2\epsilon f^2/a\,.
\end{equation}

The partially-traced propagator $\tilde{G}(p^2)$ in the momentum space  is
considered to be rederived if we take a continuum limit $\epsilon\rightarrow 0$ of 
\begin{equation}
\epsilon\sum_{t=0}^\infty N\tilde{K}_0(p^2;t)\,,
\end{equation}
which is just the discretized version of the integral (\ref{intg}), in which
$s$ corresponds to $\epsilon t$.

Fortunately, the authors of Ref.~\cite{CHJSV} have even examined the following
function:
\begin{equation}
f^c_\nu(z)\equiv\sum_{t=0}^\infty z^t I^c_\nu(t)\,,
\end{equation}
and they found the closed form of it: \cite{CHJSV}
\begin{eqnarray}
f^c_\nu(z)&=&\frac{1}{\sqrt{(1-z)^2-c^2z^2}}
\left(\frac{1-z}{cz}-\sqrt{\frac{(1-z)^2-c^2z^2}{c^2z^2}}\right)^{|\nu|}\,.
\end{eqnarray}
Using their result, we find
\begin{equation}
\sum_{q=-\infty}^\infty \cos(qN\chi)f^c_{qN}(z)=\frac{1}{\sqrt{(1-z)^2-c^2z^2}}
\frac{1-B^{2N}}{1+B^{2N}-2\cos(N\chi)B^N}\,,
\end{equation}
where 
\begin{equation}
B=B_c(z)\equiv\frac{1-z}{cz}-\sqrt{\frac{(1-z)^2-c^2z^2}{c^2z^2}}\,.
\end{equation}

A lengthy but straightforward computation gives
\begin{equation}
\epsilon\sum_{t=0}^\infty N\tilde{K}_0(p^2;t)=\epsilon N\sum_{q=-\infty}^\infty
\cos(qN\chi)f^b_{qN}(a)=\frac{N/f^2}{2\sinh\beta}
\frac{\sinh(N\beta)}{\cosh(N\beta)-\cos(N\chi)}\,,
\end{equation}
where $\frac{p^2}{f^2}=4\sinh^2\frac{\beta}{2}$. Note that $B_b(a)=e^{-\beta}$.
As the previous analysis of Ref.~\cite{KKSW2} for the flat spacetime,
this result for the propagator from the discrete time heat kernel is the same as
that from the usual continuous one, and it is even not necessary to take the limit
of $\epsilon\rightarrow 0$.

%%%%%%%%%%%%%%%%%%%%%%%%%%%%%%%%%%%%%%%%%%%%%%%%%%%%%%%%%%%%%%%%%%%%%%%%%%%
%%%%%%%%%%%%%%%%%%%%%%%%%%%%%%%%%%%%%%%%%%%%%%%%%%%%%%%%%%%%%%%%%%%%%%%%%%%
\subsection{The solution of the backward difference equation}
%%%%%%%%%%%%%%%%%%%%%%%%%%%%%%%%%%%%%%%%%%%%%%%%%%%%%%%%%%%%%%%%%%%%%%%%%%%
%%%%%%%%%%%%%%%%%%%%%%%%%%%%%%%%%%%%%%%%%%%%%%%%%%%%%%%%%%%%%%%%%%%%%%%%%%%
In this subsection, we start with the backward difference equation 
\begin{equation}
\nabla\tilde{K}_\nu(p^2;t)=\epsilon
f^2\Bigl[e^{i\chi}\tilde{K}_{\nu+1}(p^2;t)-
(2+p^2/f^2)\tilde{K}_\nu(p^2;t)+e^{-i\chi}\tilde{K}_{\nu-1}(p^2;t)\Bigr]\,.
\label{bw}
\end{equation}
Obviously, the continuum limit ($\epsilon\rightarrow 0$) of this equation is the
same differential equation (\ref{he2}) we considered.

First, following the previous subsection, we define another discrete modified
Bessel function $\bar{I}^c_\nu(t)$:
\begin{equation}
\bar{I}^c_\nu(t)\equiv\frac{(c/2)^\nu t^{\overline{\nu}}}{\nu!}F\Bigl(
\frac{\nu+t}{2},\frac{\nu+t+1}{2};\nu+1;c^2\Bigr)\,,
\end{equation}
where the rising power $t^{\overline{\nu}}$ is defined by the Pochhammer symbol,
\begin{equation}
t^{\overline{\nu}}\equiv (t)_\nu\,,
\end{equation}
which satisfies $\nabla\, t^{\overline{n}}=n\,t^{\overline{n-1}}$.
The function $\bar{I}^1_\nu(t)$ is obtained by replacing $t^k$ in 
$I_\nu(t)$ by $t^{\overline{k}}$.
The other expression of $\bar{I}^c_\nu(t)$ is
\begin{equation}
\bar{I}^c_\nu(t)=\sum_{n=0}^\infty
\frac{\Gamma(t+2n+\nu)(c/2)^{2n+\nu}}{n!\,\Gamma(t)\Gamma(\nu+n+1)}\,,
\end{equation}
and then, we see that $I^c_{-\nu}(z)=I^c_{\nu}(z)$ for $\nu\in\mathbf{Z}$.
The identity one can find is
\begin{equation}
\nabla\bar{I}^c_\nu(t)=\bar{I}^c_\nu(t)-\bar{I}^c_\nu(t-1)=
\frac{c}{2}[\bar{I}^c_{\nu+1}(t)+\bar{I}^c_{\nu-1}(t)]\,.
\end{equation}
Note also that $\bar{I}^c_\nu(0)=0$ for $\nu\ne 0$ and $\bar{I}^c_0(0)=1$. 

Now, the solution for the backward difference equation (\ref{bw}) turns out to be
\begin{equation}
\tilde{K}_\nu(p^2;t)=\sum_{q=-\infty}^\infty e^{-i(\nu+qN)\chi} \bar{a}^t
\bar{I}^{\bar{b}}_{\nu+qN}(t)\,,
\label{sumK}
\end{equation}
where
\begin{equation}
\bar{a}\equiv \Bigl[1+\epsilon f^2(2+p^2/f^2)\Bigr]^{-1}\quad\mbox{and}\quad
\bar{b}\equiv 2\epsilon f^2\bar{a}\,.
\end{equation}

As previously, we first consider the generating function
\begin{equation}
\bar{f}^c_\nu(z)\equiv\sum_{t=1}^\infty z^t \bar{I}^c_\nu(t)\,,
\end{equation}
where we should notice that the sum starts from $t=1$.
We follow the similar path as the derivation of $f^c_\nu(z)$ in Ref.~\cite{CHJSV}.
We find that the series sum
$\bar{f}^c_\nu(z)$ satisfies
\begin{equation}
(1-z)\bar{f}^c_\nu(z)=z\delta_{\nu 0}+\frac{c}{2}
(\bar{f}^c_{\nu+1}(z)+\bar{f}^c_{\nu-1}(z))\,.
\label{recf}
\end{equation}
We assume that the function takes the
form $\bar{f}^c_\nu(z)=\bar{A}_c(z)(\bar{B}_c(z))^\nu$. Then, the recursion
relation (\ref{recf}) shows
\begin{equation}
(\bar{B}_c(z))^2-\frac{2(1-z)}{c}\bar{B}_c(z)+1=0\quad\mbox{and}\quad
\bar{A}_c(z)=\frac{z}{1-z-c\bar{B}_c(z)}\,.
\end{equation}
The solution for $\bar{B}_c(z)$ is
\begin{equation}
\bar{B}_c(z)=\frac{1-z}{c}-\sqrt{\frac{(1-z)^2-c^2}{c^2}}\,,
\end{equation}
and accordingly, we obtain
\begin{eqnarray}
\bar{f}^c_\nu(z)&=&\frac{z}{\sqrt{(1-z)^2-c^2}}
\left(\frac{1-z}{c}-\sqrt{\frac{(1-z)^2-c^2}{c^2}}\right)^{|\nu|}\,.
\end{eqnarray}
This result can be verified by comparing the coefficient of the first order term of
the Maclaurin series of $\bar{f}^c_\nu(z)$ with
$\bar{I}^c_\nu(1)=\frac{1}{\sqrt{1-c^2}}
\left(\frac{c}{1+\sqrt{1-c^2}}\right)^{|\nu|}$, which is derived from the
formula in Ref.~\cite{Lebedev} for example.

Using the result above, we find
\begin{equation}
\sum_{q=-\infty}^\infty
\cos(qN\chi)\bar{f}^c_{qN}(z)=\frac{z}{\sqrt{(1-z)^2-c^2}}
\frac{1-\bar{B}^{2N}}{1+\bar{B}^{2N}-2\cos(N\chi)\bar{B}^N}\,,
\end{equation}
where $\bar{B}=\bar{B}_c(z)$,
and
\begin{equation}
\epsilon\sum_{t=1}^\infty N\tilde{K}_0(p^2;t)=\epsilon N\sum_{q=-\infty}^\infty
\cos(qN\chi)\bar{f}^{\bar{b}}_{qN}(\bar{a})=\frac{N/f^2}{2\sinh\beta}
\frac{\sinh(N\beta)}{\cosh(N\beta)-\cos(N\chi)}\,,
\end{equation}
where $\frac{p^2}{f^2}=4\sinh^2\frac{\beta}{2}$. Note that
$\bar{B}_{\bar{b}}(\bar{a})=e^{-\beta}$. This result from the discrete time heat
kernel from the backward difference equation is also the same as that from the
usual continuous one, and it is also not necessary to take the limit of
$\epsilon\rightarrow 0$.

%%%%%%%%%%%%%%%%%%%%%%%%%%%%%%%%%%%%%%%%%%%%%%%%%%%%%%%%%%%%%%%%%%%%%%%%%%%
%%%%%%%%%%%%%%%%%%%%%%%%%%%%%%%%%%%%%%%%%%%%%%%%%%%%%%%%%%%%%%%%%%%%%%%%%%%
\subsection{UV modification of the propagator}
%%%%%%%%%%%%%%%%%%%%%%%%%%%%%%%%%%%%%%%%%%%%%%%%%%%%%%%%%%%%%%%%%%%%%%%%%%%
%%%%%%%%%%%%%%%%%%%%%%%%%%%%%%%%%%%%%%%%%%%%%%%%%%%%%%%%%%%%%%%%%%%%%%%%%%%

After making the above preparation, we consider the UV modification of the
propagator.
In Ref.~\cite{KKSW2}, we introduced the modified propagator (Green's function)
of the free massive scalar field in momentum space
by omitting a finite number of discrete heat kernels $\tilde{K}(p^2;t)$, $t=0, 1,
2,\dots, n-1$ (in the notation of the present paper), in the infinite sum. 
Here, we only state the results ((2.18) in Ref.~\cite{KKSW2}),
\[\tilde{G}_n(p^2)=\frac{1}{(p^2+m^2)[1+\epsilon(p^2+m^2)]^{n-1}}\,,\]
where $m$ is the mass of the scalar field.
This
method of modification is the discrete counterpart of  the Siegel's modification
\cite{Siegel}, which converts the integration range from $[0,\infty]$ to
$[\varepsilon,\infty]$, where
$\varepsilon$ is a small constant.
It is also known that such a manner
is often used in the UV regularization in the standard heat kernel formalism.

Now, let us return to our present DD model.
For the heat kernel from the \underline{forward} difference equation, the
simplest modification method described above turns out not to be effective,
because 
$I^c_0(0)=1$, $I^c_n(t)=0$ for $t<n$ \cite{CHJSV} and
$a=1-\epsilon f^2(2+p^2/f^2)\rightarrow -\infty$ when
$p^2/f^2\rightarrow
\infty$, it is not immediately clear if the elimination of finite terms for small
numbers $t$ makes the propagator behaves better at high energy ($p^2/f^2\rightarrow
\infty$).
Another idea for the heat kernel from the forward difference equation, we
consider eliminating the infinite terms of the even powers of $a$ in the sum over
$t$. Indeed, the behavior of $\tilde{G}(p^2)$ at large $p^2$ becomes better, but
the sign change of $a$ at large $p^2$ results in a `cut' in the complex plane
of $p^2$ that are difficult to interpret.

If we adopt the heat kernel from the \underline{backward} difference equation,
the prospects of omission of finite terms seems good.
Since $\bar{a}\equiv \Bigl[1+\epsilon f^2(2+p^2/f^2)\Bigr]^{-1}>0$ for
$p^2/f^2\ge 0$, it can be easily inferred that $\tilde{K}_\nu(p^2;t)\sim
(p^2)^{-t}$ at large
$p^2$.

Therefore, we define the modified propagator in momentum space
\begin{equation}
\tilde{G}_n(p^2)\equiv\epsilon\sum_{t=n}^\infty N\tilde{K}_0(p^2;t)\,,
\end{equation}
where $\tilde{K}_0(p^2;t)$ is the solution of the \underline{backward} difference
equation. Note, of course, that $\tilde{G}_1(p^2)=\tilde{G}(p^2)$.

Here, we first study the simplest modification, the case with $n=2$.
Namely, we define
\begin{eqnarray}
& &\tilde{G}_2(p^2)=\epsilon\sum_{t=2}^\infty N\tilde{K}_0(p^2;t)=
\tilde{G}(p^2)-\epsilon N\tilde{K}_0(p^2;1)\nonumber \\
&=&\frac{N}{\sqrt{p^2(4f^2+p^2)}}\sum_{q=-\infty}^\infty\cos(qN\chi)
\left(\frac{2f^2}{2f^2+p^2+\sqrt{p^2(4f^2+p^2)}}\right)^{|q|N}\nonumber \\
&
&-\frac{N}{\sqrt{(\epsilon^{-1}+p^2)(\epsilon^{-1}+4f^2+p^2)}}\nonumber
\\
& &\times\sum_{q=-\infty}^\infty\cos(qN\chi)
\left(
\frac{2f^2}{\epsilon^{-1}+2f^2+p^2+
\sqrt{(\epsilon^{-1}+p^2)(\epsilon^{-1}+4f^2+p^2)}}\right)^{|q|N}
\,.
\end{eqnarray}

Before analyzing this modified propagator, we propose another subtraction scheme.
As the other way, we eliminate the terms of the odd order in $t$ in the
propagator. Namely, we introduce
\begin{eqnarray}
& &\tilde{G}_e(p^2)\equiv 2\epsilon\sum_{t=2,4,6\dots}^\infty
N\tilde{K}_0(p^2;t)\nonumber
\\ &=&
\frac{1}{2}(2\epsilon)\left[
N\sum_{q=-\infty}^\infty
\cos(qN\chi)\bar{f}^{\bar{b}}_{qN}(\bar{a})+
N\sum_{q=-\infty}^\infty
\cos(qN\chi)\bar{f}^{\bar{b}}_{qN}(-\bar{a})\right]\nonumber \\
&=&\frac{N}{\sqrt{p^2(4f^2+p^2)}}\sum_{q=-\infty}^\infty\cos(qN\chi)
\left(\frac{2f^2}{2f^2+p^2+\sqrt{p^2(4f^2+p^2)}}\right)^{N|q|}\nonumber \\
&
&-\frac{N}{\sqrt{(2\epsilon^{-1}+p^2)(2\epsilon^{-1}+4f^2+p^2)}}\nonumber
\\
& &\times\sum_{q=-\infty}^\infty\cos(qN\chi)
\left(
\frac{2f^2}{2\epsilon^{-1}+2f^2+p^2+
\sqrt{(2\epsilon^{-1}+p^2)(2\epsilon^{-1}+4f^2+p^2)}}\right)^{N|q|}
\,.
\end{eqnarray}

Interestingly, above two cases result in similar deformations of the
propagators:
\begin{equation}
\tilde{G}_2(p^2)=\tilde{G}(p^2)-\tilde{G}(p^2+\epsilon^{-1})\,,
\quad
\tilde{G}_e(p^2)=\tilde{G}(p^2)-\tilde{G}(p^2+2\epsilon^{-1})\,.
\end{equation}
That is, we find the same form as the propagator in the simplest Lee--Wick theory,
or almost equivalent to the one with the simplest Pauli--Villars subtraction
\cite{PV,PU,IZ,Collins,PS} as a result: For example, the propagator in momentum
space for a canonical scalar field with mass $m$ will be modified as
\begin{equation}
\frac{1}{p^2+m^2}\rightarrow\frac{1}{p^2+m^2}-\frac{1}{p^2+M^2}\,,
\end{equation}
where $M$ is the mass which would be taken as infinitely large.

The two types of modified propagators behave like $p^{-4}$ at high energies, and
the UV behavior is improved as in the Lee--Wick theory.
The important part of the results here is that both of the two subtraction
methods lead to the Lie--Wick type (albeit with two different mass parameters).
Especially in the subtraction of the odd-number terms, it is a nontrivial result
to be represented by only one parameter ($2\epsilon^{-1}$). It is also interesting
to note that, for $\tilde{G}_n~(n\ge 3)$, complicated functional forms different
from the original $\tilde{G}$ inevitably appear, and that 
$\tilde{G}_e$ is represented by being combined into a remarkable simple form.

Now, we turn to examine the behavior of divergence in $G(x,x)$ in our case for
$D=4$. If we use the cutoff scale $\Lambda$, it is written as
\begin{equation}
G(x,x)=\frac{2\pi^2}{(2\pi)^4}\int_0^\Lambda \tilde{G}(p^2)\,p^3dp\,.
\end{equation}
 If we use the modified propagators
proposed above, we can write $G(x,x)$ using the results we have obtained so far,
since 
\begin{equation}
\int_{0}^{\Lambda}\tilde{G}(p^2)\,p^3dp\rightarrow
\int_{0}^{\Lambda}
\Bigl[\tilde{G}(p^2)-\tilde{G}(p^2+M^2)\Bigr]\,p^3dp\,,
\end{equation}
where we should read $M^2=\epsilon^{-2}$ for $\tilde{G}_2$, while
$M^2=2\epsilon^{-2}$ for $\tilde{G}_e$. 
In the limit of $\Lambda\rightarrow \infty$, the only divergent term is
the term with $q=0$ in the sum-form representation of $\tilde{G}(p^2)$
(\ref{intg}).
Indeed, the contribution is found to be
\begin{eqnarray}
& &\frac{N}{16\pi^2}\Biggl[M^2\ln
\frac{\Lambda^2+M^2+2f^2+\sqrt{(\Lambda^2+M^2)(\Lambda^2+M^2+4f^2)}}{
M^2+2f^2+\sqrt{M^2(M^2+4f^2)}}+
\sqrt{M^2(M^2+4f^2)}\nonumber \\
& &-
2f^2\ln\frac{M^2+2f^2+\sqrt{M^2(M^2+4f^2)}}{2f^2}-\frac{M^2(2\Lambda^2+M^2+4f^2)}{\sqrt{(\Lambda^2+M^2)(\Lambda^2+M^2+4f^2)}+
\sqrt{\Lambda^2(\Lambda^2+4f^2)}}\nonumber \\
& &+2f^2\ln
\frac{\Lambda^2+M^2+2f^2+\sqrt{(\Lambda^2+M^2)(\Lambda^2+M^2+4f^2)}}{
\Lambda^2+2f^2+\sqrt{\Lambda^2(\Lambda^2+4f^2)}}
\Biggr]\,,
\label{g2}
\end{eqnarray}
and this gives the logarithmic divergence $\sim M^2\ln\Lambda$ when
$\Lambda\rightarrow\infty$, instead of the quadratic divergence $\sim \Lambda^2$
known in the ordinary scalar one-loop effect.

Now we consider $\tilde{G}_3(p^2)$ and the diverging behavior of $G_3(x,x)$.
First we note that
\begin{equation}
\tilde{G}_3(p^2)=\tilde{G}_2(p^2)-\epsilon N\tilde{K}_0(p^2;2)\,,
\end{equation}
and the calculations on the divergent part above can be used.
Since $\bar{I}_0^c(2)=\frac{1}{(1-c^2)^{3/2}}$, the divergent contribution comes
from the $q=0$ term in the sum form of $\epsilon N\tilde{K}_0(p^2;2)$ (\ref{sumK})
and it turns out to be
\begin{equation}
\frac{N\epsilon^{-1}(p^2+2f^2+\epsilon^{-1})}{
[(p^2+\epsilon^{-1})(p^2+4f^2+\epsilon^{-1})]^{3/2}}\,.
\label{frac}
\end{equation}
Note that this behaves $\sim 1/p^4$ for large $p$.
One can find that the momentum integration of $\epsilon
N\tilde{K}_0(p^2;2)$ (\ref{frac}) with the cutoff $\Lambda$ is 
\begin{eqnarray}
& &\frac{N}{16\pi^2}\Biggl[M^2\ln
\frac{\Lambda^2+M^2+2f^2+\sqrt{(\Lambda^2+M^2)(\Lambda^2+M^2+4f^2)}}{
M^2+2f^2+\sqrt{M^2(M^2+4f^2)}}\nonumber \\
& &-\frac{M^2\Lambda^2}{\sqrt{(\Lambda^2+M^2)(\Lambda^2+M^2+4f^2)}}\Biggr]\,,
\label{g3}
\end{eqnarray}
where $M^2=\epsilon^{-1}$.
Then, subtraction of (\ref{g3}) from (\ref{g2}) gives
\begin{eqnarray}
G_3(x,x)&=&\frac{N}{16\pi^2}\Biggl[
\sqrt{M^2(M^2+4f^2)}-
2f^2\ln\frac{M^2+2f^2+\sqrt{M^2(M^2+4f^2)}}{2f^2}
\Biggr]\nonumber \\
& &+(\mbox{finite, $\chi$-dependent terms})\,,
\end{eqnarray}
in the lmit of $\Lambda\rightarrow\infty$.
The result that $G_3(x,x)\sim M^2$ is an expected one, but is involving the scale
$f$ in a nontrivial form.

%%%%%%%%%%%%%%%%%%%%%%%%%%%%%%%%%%%%%%%%%%%%%%%%%%%%%%%%%%%%%%%%%%%%%%%%%%%
%%%%%%%%%%%%%%%%%%%%%%%%%%%%%%%%%%%%%%%%%%%%%%%%%%%%%%%%%%%%%%%%%%%%%%%%%%%
%%%%%%%%%%%%%%%%%%%%%%%%%%%%%%%%%%%%%%%%%%%%%%%%%%%%%%%%%%%%%%%%%%%%%%%%%%%
\section{Summary and outlook}
\label{conclusion}
%%%%%%%%%%%%%%%%%%%%%%%%%%%%%%%%%%%%%%%%%%%%%%%%%%%%%%%%%%%%%%%%%%%%%%%%%%%
%%%%%%%%%%%%%%%%%%%%%%%%%%%%%%%%%%%%%%%%%%%%%%%%%%%%%%%%%%%%%%%%%%%%%%%%%%%
%%%%%%%%%%%%%%%%%%%%%%%%%%%%%%%%%%%%%%%%%%%%%%%%%%%%%%%%%%%%%%%%%%%%%%%%%%%

In the present paper, we consider the discrete time heat kernel in the simple
model of the dimensionally deconstructed scalar QED. The Euclidean propagator of
the scalar field is obtained by the sum of the discretized kernels, which are not
individually expressed by the direct products of the kernel on the flat space and
that on the discrete circle. The propagator thus obtained by an infinite sum of
kernels is found to be in perfect agreement with the usual one obtained from the
integral of the continuous time heat kernel. A nontrivial result here is that it
holds in both forward and backward discretization cases, and not relying on the
scale of the discretization unit
$\epsilon$, that is, there is no need to take the limit of $\epsilon\rightarrow 0$.

Furthermore, we considered the behavior of the two-point coincidence limit of the
propagator $G(x,x)$, which is a measure of the UV divergence of the one-loop mass
correction to the interacting scalar field. For a usual propagator, it gives a
quadratic divergence in the four-dimensional spacetime, namely,
$G(x,x)\sim\Lambda^2$, where $\Lambda$ is the cutoff momentum.
In the previous paper \cite{KKSW2}, we proposed the method to modify the
propagator by subtracting contributions of a finite number of discretized heat
kernels. This idea is a discretized version of Siegel's method \cite{Siegel}
of moving the lower bound of the integration range of the kernel.
We apply similar method to the present model with DD.
We found that both cases with the subtraction of the first kernel of $t=1$
and with the subtraction of the odd-numbered kernels of $t=1,3,5,\dots$ gives
the same form of the resulting propagator, $G(p^2)-G(p^2+M^2)$; $M^2=\epsilon^{-1}$
in the former and $M^2=2\epsilon^{-1}$. This is a nontrivial result, and
this modified propagator has exactly the same structure as that of the Lee--Wick
theory, or the regularized propagator of Pauli--Villars. Consequently, this
modified propagator gives $G(x,x)\sim \ln\Lambda$ in four dimensions.
Finally, we showed that the modified propagator by subtracting the kernels of
$t=1$ and $t=2$ gives $G(x,x)\sim M^2=\epsilon^{-1}$.

%%%%%%%%%%%%%%%%%%%%%%%%%%%%%%%%%%%%%%%%%%%%%%%%%%%%%%%%%%%%%%%%%%%%%%%%%%%

It turns out that the exact propagator can be obtained without taking the
continuum limit with the solutions of two difference equations. From this fact,
it seems that we may be able to develop more free ideas regarding the handling of
the proper time method, even though it looks like an ad hoc assumption at the
current primitive stage of study.
Of course, it is necessary to pursue the principle underlying the physical
inevitability of use of discrete proper-time.

Anyway, the discretization of the heat kernel for the Dirac operator on
continuous and discrete real spaces is interesting mathematically
and may be important for physical applications such as an approach to the hierarchy
problem. Since the Dirac operators on graphs  have been studied so far (see for
example, Ref.~\cite{KS2}), the study on heat kernel for them can be tackled in
near future.

Although it may seem trivial in the present treatment, it is interesting that the
finite part involving the PNGB field is also subject to change when the propagator
is modified under our prescription. This effect would similarly affect the
Hosotani-like mechanism in DD theory, though the treatment of the divergence in
the potential requires further consideration.

The UV modified propagator obtained in our method includes additional mass poles in
the complex plane of $-p^2$. Such poles appear in general higher derivative
theories and have been studied by many authors.
In the higher derivative theory, the addition of interaction may bring about specific instability and
may violate unitarity in the Lorentzian spacetime, since analysis with auxiliary
fields indicates the appearance of negative norms. This is a subject that has
been frequently addressed and has recently been actively discussed
\cite{AP,Anselmi1,Anselmi2}.
Unfortunately, our current understanding has not led us to discuss the unitarity
of the models presented in this paper in more detail.
We consider that detailed discussions about unitarity and other
physics on self-interacting models should be studied further. At the same time, we
have to reconsider the heat kernel derived from the forward difference equation,
which yields the momentum-space propagator involving an apparently bizarre cut when
discrete subtraction is carried out, in similar sense.%
\footnote{It is likely to be related to the studies on field theories with not only
higher derivatives but also fractional powers of derivative operators \cite{CCLR}.}

Finally, we would like to write about the discretization of real whole spacetime,
which is a radical extension of the DD model, and the study of its heat kernel as
well. Mathematically, it would be straightforward to replace the continuum
space with a certain graph structure.
Many difficulties can be expected from physical requirements, but it will be great
if we could use the mathematics of heat kernel theory to uncover non-obvious
features of the general discretized models.

%%%%%%%%%%%%%%%%%%%%%%%%%%%%%%%%%%%%%%%%%%%%%%%%%%%%%%%%%%%%%%%%%
%%%%%%%%%%%%%%%%%%%%%%%%%%%%%%%%%%%%%%%%%%%%%%%%%%%%%%%%%%%%%%%%%
%\appendix
%%%%%%%%%%%%%%%%%%%%%%%%%%%%%%%%%%%%%%%%%%%%%%%%%%%%%%%%%%%%%%%%%

%%%%%%%%%%%%%%%%%%%%%%%%%%%%%%%%%%%%%%%%%%%%%%%%%%%%%%%%%%%%%%%%%%%%%%%%%%%
%\acknowledgments
%%%%%%%%%%%%%%%%%%%%%%%%%%%%%%%%%%%%%%%%%%%%%%%%%%%%%%%%%%%%%%%%%%%%%%%%%%%
%Acknowledgements
%%%%%%%%%%%%%%%%%%%%%%%%%%%%%%%%%%%%%%%%%%%%%%%%%%%%%%%%%%%%%%%%%%%%%%%%%%%
%\begin{acknowledgments}
%The authors are grateful to Ryo Saito for useful discussions.
%the organizers of JGRG21, where our
%partial result %({\tt [arXiv:10mm.xxxx]}) 
%was presented. %for elucidating comments.
%This study is supported in part by the Grant-in-Aid of Nikaido Research 
%Fund.
%\end{acknowledgments}
%%%%%%%%%%%%%%%%%%%%%%%%%%%%%%%%%%%%%%%%%%%%%%%%%%%%%%%%%%%%%%%%%%%%%%%%%%%

%%%%%%%%%%%%%%%%%%%%%%%%%%%%%%%%%%%%%%%%%
%%%%%%%%%%%%%%%%%%%%%%%%%%%%%%%%%%%%%%%%%
%%%
%%% References
%%%
%%%%%%%%%%%%%%%%%%%%%%%%%%%%%%%%%%%%%%%%%
%%%%%%%%%%%%%%%%%%%%%%%%%%%%%%%%%%%%%%%%%
%%%%%%%%%%%%%%%%%%%%%%%%%%%%%%%%%%%%%%%%%%%%%%%%%%%%%%%%%%%%%%%%%%%%%%%%%%%
%thebibliography
%%%%%%%%%%%%%%%%%%%%%%%%%%%%%%%%%%%%%%%%%%%%%%%%%%%%%%%%%%%%%%%%%%%%%%%%%%%
%\bibliographystyle{apsrev}
\bibliographystyle{apsrev4-1}
%\bibliography{}

%%%%%%%%%%%%%%%%%%%%%%%%%%%%%%%%%%%%%%%%%%%%%%%%%%%%%%%%%%%%%%%%%%%%%%%%%%%
%%%%%%%%%%%%%%%%%%%%%%%%%%%%%%%%%%%%%%%%%%%%%%%%%%%%%%%%%%%%%%%%%%%%%%%%%%%
%%%%%%%%%%%%%%%%%%%%%%%%%%%%%%%%%%%%%%%%%%%%%%%%%%%%%%%%%%%%%%%%%%%%%%%%%%%
\end{document}